# Self-Contained, Cooled SiPM Array for Scintillation Spectroscopy

A. Ponento[a], C. J. Martoff[a], D. Jones[a] and E. Kaczanowicz[a]

[a] Temple University,
 1925 N 12th St, Philadelphia, PA, 19122, USA
 E-mail: cmartoff@protonmail.com

ABSTRACT: A simple, self-contained, thermo-electrically cooled SiPM system is presented which cools a SiPM array to -20° C. The array views a NaI scintillator through a 75 mm diameter glass window. Waste heat is removed with a large heat sink and AC fans. Above 40 keV in an air-coupled 2" x 2" NaI scintillator, the SiPM dark count rate was reduced by a factor ~1000 when cooled. Performance when cooled was very similar to a PMT tested in the same setup and adequate for nuclear spectroscopy above 25 keV. Originally, water cooling was used but it was replaced by air cooling which is more suitable for a self-contained system, giving the advantage of portability without degrading performance. Straightforward improvements would allow cooling the SiPM to -30° C or below, which would further reduce the dark count rate and extend the spectroscopically useful range to even lower energies. Such a system would be rugged and suitable for field use, for instance for inspection of cargo for gamma ray emissions.

KEYWORDS: Gamma detectors; Spectrometers; Detector cooling and thermo-stabilization; Detector design and construction technologies and materials.

**Contents**

**1. Introduction**

Photomultiplier tubes (PMT) are still the workhorses of low level (including single-photon) light detection. These vacuum tube devices work by electron-impact multiplication of photo-electrons. They are widely used in spectroscopy, particle physics, astrophysics, medical diagnosis, and military technology. However, a solid-state replacement for the PMT, called the Silicon Photomultiplier (SiPM) has been developed and is increasing in popularity. SiPMs have numerous advantages over PMTs; they are much smaller, more rugged, less sensitive to stray magnetic fields, consume less power, require no kilovolt bias, and are less expensive in some applications.

Unfortunately, the major drawback of SiPMs is their high dark current. Dark current, sometimes referred to as dark noise, is the signal produced in a light detector in the absence of any incident light. At room temperature, SiPMs have dark current from thermally excited electrons equivalent to tens or hundreds of thousands of electrons per second per mm2.[1] This high noise level prevents spectroscopy down to the level of few-photon pulses, usually completely prevents large-area single photoelectron detection, and significantly degrades the spectroscopic energy resolution for larger pulses containing hundreds or thousands of photoelectrons, like those typically examined using nuclear physics scintillators used to identify radioactive substances.

The purpose of the present work is to show a compact, self-contained SiPM-based scintillation detector in which the SiPM dark noise is drastically reduced by cooling the SiPM to -20° C with thermoelectric coolers (TECs). The considerable waste heat (170 Watts) from the TECs is removed by forced air cooling. This is achieved in a small and self-contained setup.

---

[1] "C Series Datasheet." Sensl, Apr. 2018.

## 2. Silicon Photomultiplier

2.1 Model

The SiPM that was a Sensl ArrayB-30035-144P.[2] This is a 50 by 50 mm2 mosaic of 144 pixels, each 3 millimeters on a side. Each pixel is an array of 35 by 35 micron avalanche photodiodes (APDs), giving an overall 72% active area fraction. These sensors detect light by summing the saturated outputs from all avalanche photodiodes (APDs) in a pixel. This model was selected to approximate a 2" PMT. Readout was accomplished using the tileable AB424T-ARRAY144P board from AiT Industries.[3] This flexible board has a diode-coupled front-end to reduce noise, and offers single element, row-and-column, and position-encoded readout through the SiPMIM4 interface module, as well as bias voltage and SiPM temperature and current readouts.

2.2 Dark Count

At room temperature, Sensl specifies the dark current of the 3 mm pixels to be 2.8 µA at typical levels.[4] With the specified gain of $3x10^6$, this translates into a dark count of roughly 16 MHz per pixel, much too high to detect any low-level light signals. Single photon detection with SiPMs can be performed with small area devices and a tight time coincidence, either between devices on a single scintillator fiber[5] or with an external time mark. However, for straightforward nuclear spectroscopy applications with large collecting area, the dark current represents a serious limitation. Since the dark current results from thermally activated carriers in the silicon,cooling the solid stated device should significantly reduce the problem. Typically, SiPM dark count decreases by an order of magnitude for each 30K decrease in temperature, until about 90K where the noise remains constant at approximately $1x10^{-2}$ Hz/mm2.[6]

---

[2] "ArrayB-300Series-144P Preliminary Datasheet." Sensl, May 2013.
[3] "AB424T-ARRAY144P." AiT Instruments. N.d.
[4] "B-Series Preliminary Datasheet." Sensl, Feb. 2013.
[5] Moutinho, L.M., et al. "Development of a Scintillating Optical Fiber Dosimeter with Silicon Photomultipliers." Nuclear Instruments and Methods in Physics Research Section A: Accelerators, Spectrometers, Detectors and Associated Equipment, 2011, doi:10.1016/j.nima.2011.11.069.
[6] Acerbi, Fabio, et al. "Cryogenic Characterization of FBK HD Near-UV Sensitive SiPMs." IEEE Transactions on Electron Devices, vol. 64, no. 2, 2017, pp. 521–526., doi:10.1109/ted.2016.2641586.

## 3. Experimental Apparatus

### 3.1 SiPM Enclosure

If cooled below the dew point of the laboratory (around 10° C for typical indoor environments), cooled parts must be isolated from the laboratory atmosphere to avoid condensation. This can be done with an enclosure containing dry gas or vacuum. The thermal conductivity of gases is nearly independent of pressure down to below one mTorr, so the choice of isolation hardly affects the thermal performance unless rather low pressures can be achieved. In the present work we used a rough vacuum of about 500 milliTorr.

Inside the rough vacuum, the SiPM was placed inside an open copper frame, which makes pressure contact with the cold side of the thermoelectric cooler. The open side of the copper frame held a sapphire window mounted with thermal coupling compound, and the SiPM was mounted against the window with springs and optical grease, providing the thermal contact with the TEC. Sapphire was chosen because of its excellent optical viewing characteristics, strength and stability, and high thermal conductivity.

The vacuum was contained within a stainless steel cylinder chamber. A rotary vacuum pump connected to its base maintained the rough vacuum, and the interior electronics received power via feedthroughs. A borosilicate glass window in front of the sapphire enabled the SiPM assembly to be air-coupled to a typical encapsulated NaI scintillator.

### 3.2 Thermoelectric Cooler

Cooling was accomplished using thermoelectric coolers (TECs), Marlow model XLT2422[7], which are solid-state devices that transport heat against a temperature gradient via the Peltier Effect.

The present realization uses a single "primary" TEC inside the enclosure, in indirect contact with the SiPM. The TEC specification gives a maximum temperature difference between hot and cold sides of 66° C. Thus, to achieve a SiPM temperature of 20° C requires that the hot side of the primary TEC be at most 25° C. This temperature is directly determined by the efficiency with which heat (both the radiation and conduction heat load of the system and the Joule heating from the TEC power input) is removed from the hot side of this inner TEC. To achieve the present level of performance, four additional TECs mounted in parallel, outside the enclosure were found to be required to maintain the inner TEC hot side temperature below 30° C. Finally, the heat of the four outside TECs is removed by forced air cooling with a large heat sink in the room air.

### 3.3 Estimated Heat Loads

| Heat Load (W) | Section I | Section II | Section III |
|---|---|---|---|
| Conductive | 1.6 | -- | -- |
| Convective | 1.1 | 0.8 | -- |
| Radiative | 2.9 | 0.8 | -- |
| Electric | 0.22 | 33 | 170 |
| Total | 5.8 | 35 | 170 |

**Table 1.** Estimated heat loads for present design.

---

[7] "Technical Data Sheet for XLT2422." Marlow Industries. N.d.

Table 1 shows the heat loads of the system, where the "sections" are indicated in Figure 1. Values for the conduction, convection, and radiation heat loads on each section have been evaluated using standard formulae.[8] Section I is the coldest stage and includes the SiPM. Heat load is calculated from the surrounding stainless steel vacuum chamber and borosilicate window at about 21° C to the copper casing and sapphire window at about -20° C, respectively. Section II is between the single TEC and the four parallel TECs, where conduction is negligible. Convection is calculated by considering the heat flow from the surrounding air at 23.5° to the large copper piece mounted on the vacuum chamber at about 17° C. The radiation is calculated as the heat flow from the surrounding walls to the same copper piece. The electric heat load is calculated from the electric current used by the TECs. Section III is the final heat load transferred to the heat sink. There are no contributions to the heat load from conduction, convection, or radiation as the temperature of it (50° C) is much hotter than the surrounding air (23.5° C). The electric section is determined in the same manner as in Section II. This section includes the heat sink and thus handles the final total heat waste of 170W through forced convection. The heat load of the system for Sections II and III is mainly dominated by the Joule heating of the electric power supplied to the solid-state coolers. While radiation, conduction, convection, and to a lesser extent electric power of the SiPM heat up the device in Section I, their overall contribution across all sections is miniscule compared to that of the current powering the solid-state coolers. These require about 3A and 17V DC each, and there are five in our device. This creates a total waste heat load of approximately 170W.

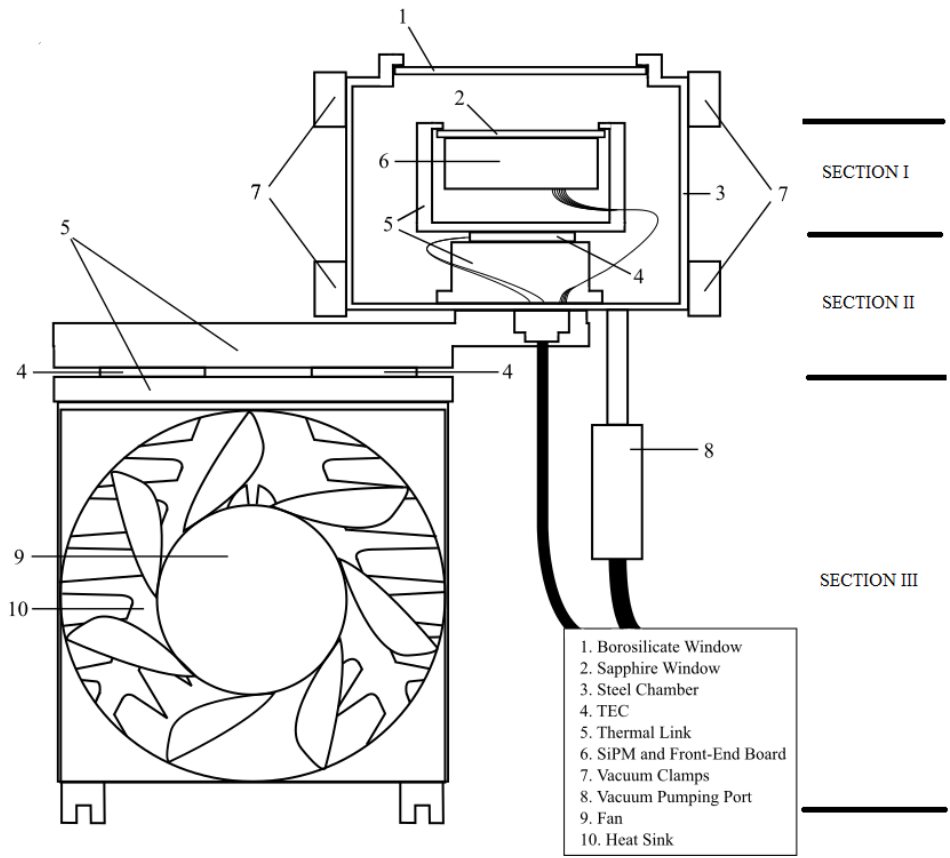

**Figure 1.** Side view of the cooling system designed for the SiPM. The thermal sections referred to in Table 1 are indicated at right.

---

[8] Chhabra, R. P. *CRC Handbook of Thermal Engineering*. 2nd ed., Taylor & Francis, CRC Press, 2017.

3.4 Heat Sink

To remove the 170 Watts of waste heat, a Wakefield-Vette I392-120AB[9] heat sink is attached to the hot side of four external TECs through a copper plate, as configured in Figure 2. This sink is aluminum, with a specified thermal resistance of only 0.11° C/Watt, and measures approximately 120 by 120 by 130 mm. To achieve the specified thermal resistance and the desired copper plate temperature of 50° C, we used forced air cooling with a push-pull set of two Orion AC powered muffin fans, model OA109AP-11-TBXC.[10] This heat sink system maintained the desired external copper plate at a temperature of 50° C, which resulted in a measured SiPM temperature of -20° C as measured through the AiT interface board.

Initial trials were also performed with the external TECs cooled by flowing water at 1.5 gpm, using a water-cooled heat sink. This resulted in similar performance to the air-cooled system, without the advantage of portability.

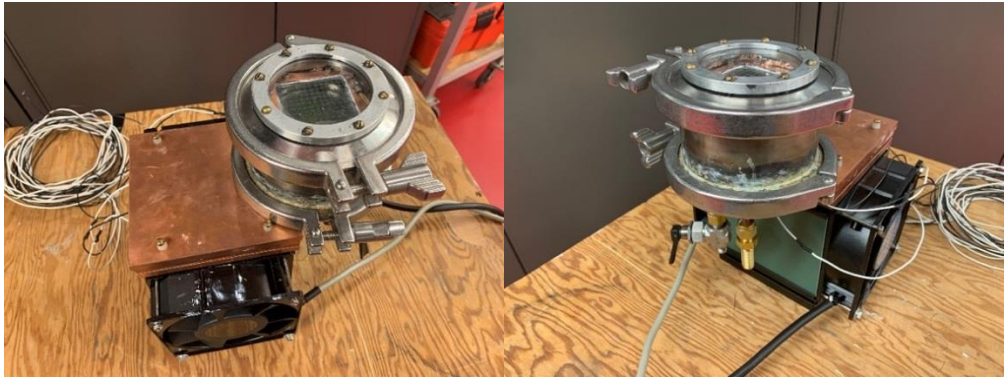

**Figure 2.** On the left, the present device shown with the borosilicate viewing window at top right and the SiPM array visible below it. Limited space on the lower vacuum flange required the copper heat path to be offset as shown. While this interfered with efficient removal of heat by the external TECs, the offset would be absent in a more advanced design. One of the fans is visible as the black object at left front. On the right, one of the fiberglass plates confining airflow around the heatsink, and also the second fan, are shown underneath the copper plate.

## 4. Experimental Results

4.1 Dark Count Measurements

| Temperature (°C) | 25 | -20 |
|---|---|---|
| Bias (V) | 32 | 32 |
| Threshold (mV) | Dark Count (kHz) | Dark Count (kHz) |
| 25 | 6630 | 190 |
| 50 | 4670 | 1 |
| 75 | 2410 | 0.08 |
| 100 | 1070 | 0.04 |
| 125 | 360 | 0.01 |

**Table 2.** Dark Count of SiPM as a function of threshold and temperature.

---

[9] "Wakefield-Vette 392 Series." Wakefield-Vette. N.d.
[10] "Orion Fans OA109 'XC Series.'" Knight Electronics, Inc. N.d.

Table 2 shows data for the dark count of the SiPM as a function of temperature and threshold. The rates in Table II were obtained using a discriminator counting pulses from the ORTEC 673 spectroscopy amplifier used to obtain the pulse height spectra. As seen in Table 2, the dark count is decreased by factors ranging up to 100,000 at the highest thresholds. As seen in the next section, this results in a significant improvement in spectroscopic performance at -20° C compared to data without cooling. The sharp rise in SiPM background occurs close to the low energy peaks around 25 keV in the pulse height spectra. Considering the relatively poor light coupling in the present setup, we estimate this to correspond to less than 100 photoelectrons.

4.2 Cs-137

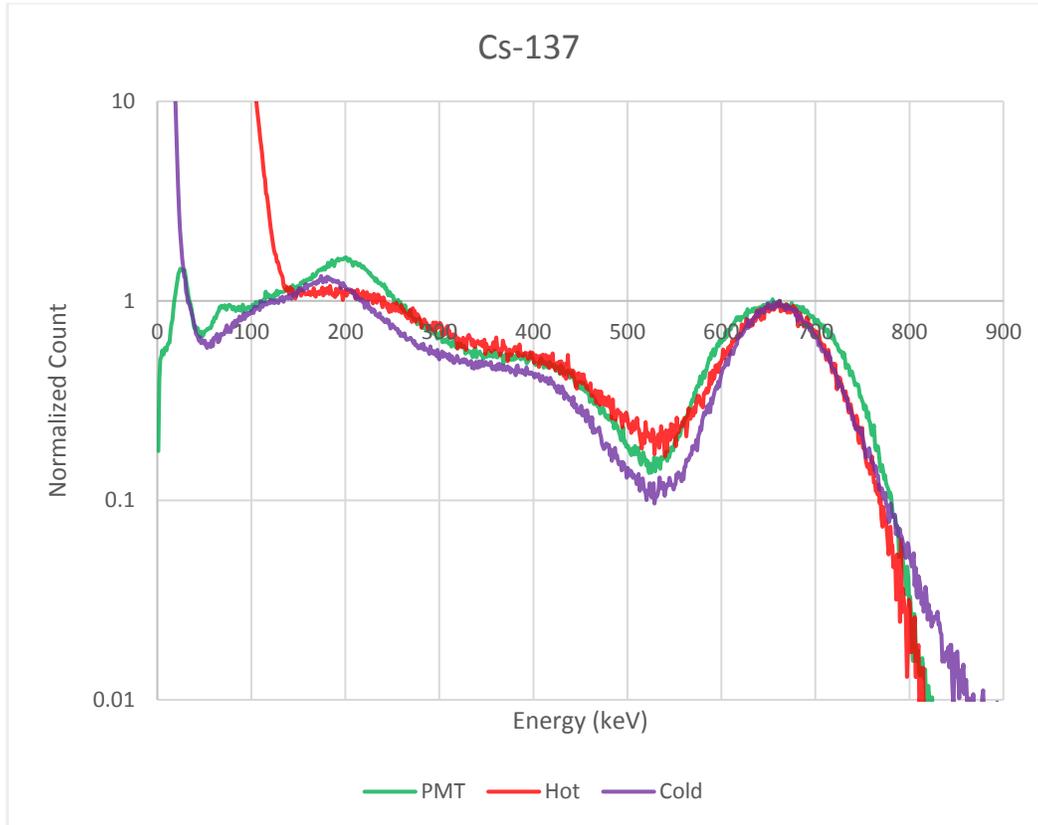

**Figure 3.** Pulse height spectra obtained with ORTEC 673 Spectroscopy Amplifier and Gated Integrator and Amptek MCA-8000A from the present device viewing 2 x 2 NaI exposed to Cs-137 source. Spectra features are discussed in the text.

Nuclear spectroscopy tests were performed with the array using a 2x2" NaI scintillator and gamma ray sources. The NaI was placed in contact with the viewing window of the SiPM apparatus without use of optical coupling grease. The SiPM was connected to an AiT SIPMIM4-BNC interface, which was then connected to an ORTEC 673 Spectroscopy Amplifier and Gated Integrator. This was then interpreted by an Amptek MCA-8000A multichannel analyzer and viewed on a desktop by the program ADMCA. For comparison, a Photonis XP2230 PMT (2" diameter head-on borosilicate glass) was alternately used to view the same NaI. The first source used was Cs-137, which has a single gamma ray peak at 662 keV. The Cs-137 Compton edge occurs around 450 keV, a Compton plateau from around 300 to 400 keV, and a backscatter peak at approximately 200 keV. The energy scale was assigned based on the known photopeak energies of the 137-Cs and 133-Ba sources using a linear calibration fit.

In Figure 3, the normalized data for three measurements is shown: SiPM at room temperature, SiPM cooled, and room temperature PMT. The channels were normalized around the peak energy, 662 keV, and the counts were normalized by dividing all counts by the count at the peak. All spectra resolve the photopeak, but the photopeak FWHM ratio was 20% at 25° C and 16% at -20° C. Another resolution measure is the valley to peak ratio between the Compton edge (525 keV) and the photopeak. These are 18% at 25 degrees and 9.6% at -19 degrees. Qualitatively, one can also see the Compton edge, Compton plateau, and backscatter peak in the cold SiPM data while only the Compton edge is slightly visible in the room temperature SiPM data. The PMT has superior noise performance only below about 25 keV.

4.3 Ba-133

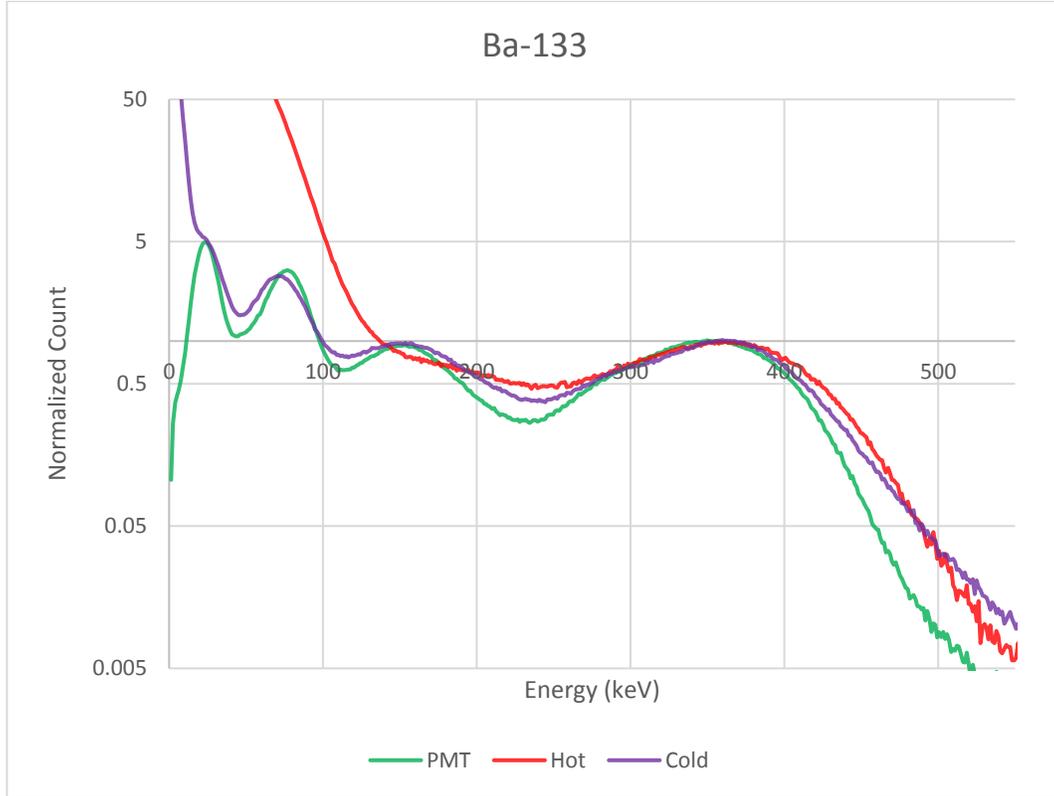

**Figure 4.** Pulse height spectra obtained with ORTEC 673 Spectroscopy Amplifier and Gated Integrator and Amptek MCA-8000A from the present device viewing 2 x 2 NaI exposed to Ba-133 source.

The low energy performance is more clearly seen performing the same tests with a Ba-133 source, which emits gamma rays at 356, 303, 81, and 31 keV.

As seen in Figure 4, the SiPM resolves only a broad maximum in the region of the 356 keV peak at room temperature. When cooled, the spectrum shows that this peak is made up of several gamma ray emissions; namely, the 356 and 303 keV begin to be resolved when cooled. In addition, the expected low energy peaks at 81, and even 31 keV (a shoulder) are visible only in the cooled SiPM (and PMT) spectra. Generally, the SiPM shows all the same peaks as the PMT measurements, with just slightly worse valley to peak ratios. With the present level of cooling, dark noise still obscures the spectrum below approximately 25 keV.

## 5. Conclusions

Nuclear spectroscopy testing shows that the present SiPM cooled to -20° C performs significantly better than at room temperature. With non-optimized optical coupling, the cooled SiPM performance is nearly identical to that of a PMT for energies above 25 keV in NaI. The dark noise is drastically reduced by cooling. This has been accomplished in a portable, air-cooled system, maintaining the various advantages that the SiPM has over the PMT for handheld and/or rugged-conditions applications.

The present isolation and cooling system is considered only a proof-of-principle. For a widely deployable nuclear spectroscopy system, aspects of the design require further improvement. Similar results were obtained with water cooling, but this compromised our goal of a self-contained system.

The overall mechanical design of the system would be improved using a smaller chamber with improved flanges. The need for a vacuum pump would be eliminated by use of a sealed, gas filled system. It would alternatively be possible to eliminate the conduction and convection heat load by using a sealed high vacuum enclosure with integral getter pump, reducing Section I heat load by 50%. Additional superinsulation could also be employed to further reduce the radiation heat load on the copper holder. Heat flow calculations show that the present copper holder should be streamlined, and the number of thermal junctions reduced to improve the heat flow. These improvements would result in even lower SiPM temperatures with the present array of TECs.

Multi-stage TECs are available for the primary cooler, with much greater hot to cold side temperature differences, but these cannot handle as much heat. With a re-design to incorporate one or more of these, the SiPM could be cooled to substantially lower temperatures and achieve even better spectroscopic performance.

**Acknowledgements**

This work was supported by Temple University and its Science Scholars Program. We would like to thank machinist Matthew McCormick and electronics engineer Richard Harris for valuable assistance in the realization of the device. In addition, we want to thank Steven Sanderlin for making important contributions to the initial thermal design.